\documentclass[11pt]{article}
\usepackage{amsfonts,latexsym}
\usepackage{amsmath}
\usepackage{amscd}
\usepackage{float,amsmath,amssymb,mathrsfs,bm,multirow,graphics}
\usepackage[dvips]{graphicx}
\usepackage[all]{xy}

\newcommand{\nequation}{\setcounter{equation}{0}}

\newcommand{\R}{{\Bbb R}}

\newcommand{\C}{{\Bbb C}}
\newcommand{\Z}{{\Bbb Z}}

\newcommand{\umin}{{z}}
\newcommand{\umax}{{Z}}

\newtheorem{theorem}{Theorem}[section]

\newtheorem{remark}[theorem]{Remark}

\newtheorem{figuretext}{Figure}

\begin{document}

\thispagestyle{empty}
\vspace{2cm}
\begin{flushright}
 ITP-UH-17/08  
\\[1cm]
\end{flushright}
\begin{center}
{\Large\bf On the $N{=}2$ Supersymmetric Camassa-Holm and Hunter-Saxton Equations}
\end{center}
\vspace{1cm}

\begin{center}
{\large\bf
J. Lenells${\,}^{a}$ \ and \
O. Lechtenfeld${\,}^{b}$}
\end{center}

\begin{center}

\vspace{0.2cm}
${}^a$ {\it
Department of Applied Mathematics and Theoretical Physics, 
	\\
University of Cambridge, Cambridge CB3 0WA, United Kingdom}
\vspace{0.2cm}

${}^b$ {\it
Institut f\"ur Theoretische Physik, Leibniz Universit\"at Hannover,} \\
{\it Appelstra\ss{}e 2, 30167 Hannover, Germany}
\vspace{0.2cm}

{\tt j.lenells@damtp.cam.ac.uk,  lechtenf@itp.uni-hannover.de}
\end{center}
\vspace{2cm}

\begin{abstract}
\noindent
We consider $N{=}2$ supersymmetric extensions of the Camassa-Holm and Hunter-Saxton equations.  We show that they admit geometric interpretations as Euler equations on the superconformal algebra of contact vector fields on the $1|2$-dimensional supercircle. We use the bi-Hamiltonian formulation to derive Lax pairs. Moreover, we present some simple examples of explicit solutions. As a by-product of our analysis we obtain a description of the bounded traveling-wave solutions for the two-component Hunter-Saxton equation. 
\end{abstract}
\vfill

\newpage
\setcounter{page}{1}

\section{Introduction} \nequation
The Camassa-Holm (CH) equation 
\begin{equation}\label{CH}\tag{CH}
               u_t - u_{txx} + 3uu_x  = 2u_xu_{xx}+uu_{xxx}, \qquad x\in\R,\;t>0,
\end{equation}
and the Hunter-Saxton (HS) equation 
\begin{equation}\label{HS}\tag{HS}
   u_{txx} = - 2u_xu_{xx}-uu_{xxx}, \qquad x\in\R,\;t>0,
\end{equation}
where $u(x,t)$ is a real-valued function, are integrable models for the propagation of nonlinear waves in $1 + 1$-dimension. 
Equation (\ref{CH}) models the propagation of shallow water
waves over a flat bottom, $u(x,t)$ representing the water's free surface
in non-dimensional variables. It was first obtained mathematically \cite{F-F} as an abstract equation with two distinct, but compatible, Hamiltonian formulations, and
was subsequently derived from physical principles \cite{C-H, C-Lannes, F-Liu, J1}. Among its most notable properties is the existence of peaked solitons \cite{C-H}. 
Equation (\ref{HS}) describes the evolution of nonlinear oritentation waves in liquid crystals, $u(x,t)$ being related to the deviation of the average orientation of the molecules from an equilibrium position \cite{H-S}. 

Both (\ref{CH}) and (\ref{HS}) are completely integrable systems with an infinite number of conservation laws (see e.g. \cite{C-E0, C-K2, C-M, GH, H-Z, K-M}).
Moreover, both equations admit geometric interpretations as Euler equations for geodesic flow on the diffeomorphism group $\text{Diff}(S^1)$ of orientation-preserving diffeomorphisms 
of the unit circle $S^1$. More precisely, the geodesic motion on $\text{Diff}(S^1)$ endowed with the right-invariant metric given at the identity by
\begin{equation}\label{H1metric}
  \langle u, v \rangle_{H_1} = \int_{S^1} \bigl(uv + u_x v_x\bigr) dx,
\end{equation}
is described by the Camassa-Holm equation \cite{Mis98} (see also \cite{CKKT, C-K}), whereas (\ref{HS}) describes the geodesic flow 
on the quotient space $\text{Diff}(S^1)/S^1$ equipped with the right-invariant metric given at the identity by \cite{K-M}
\begin{equation}\label{H1dotmetric}
\langle u, v \rangle_{\dot{H}_1} = \int_{S^1} u_x v_x dx.
\end{equation}

We will consider an $N =2$ supersymmetric generalization of equations (\ref{CH}) and (\ref{HS}), which was first introduced in \cite{P2006} by means of bi-Hamiltonian considerations. In this paper we: (a) Show that this supersymmetric generalization admits a geometric interpretation as an Euler equation on the superconformal algebra of contact vector fields on the $1|2$-dimensional supercircle;\footnote{
The interpretation of the $N=2$ supersymmetric Camassa-Holm  
equation as an Euler equation related to the superconformal algebra was  
already described in \cite{AGZ}.}
(b) Consider the bi-Hamiltonian structure; (c) Use the bi-Hamiltonian formulation to derive a Lax pair; (d) Present some simple examples of explicit solutions.

\subsection{The supersymmetric equation}
In order to simultaneously consider supersymmetric generalizations of both the CH and the HS equation, it is convenient to introduce the following notation:
\begin{itemize}
\item $\gamma \in \{0,1\}$ is a parameter which satisfies $\gamma = 1$ in the case of CH and $\gamma = 0$ in the case of HS.

\item $\Lambda = \gamma - \partial_x^2$.

\item $m = \Lambda u$.

\item $\theta_1$ and $\theta_2$ are anticommuting variables.

\item $D_j = \partial_{\theta_j} + \theta_j \partial_x$ for $j = 1,2$.

\item $U = u + \theta_1 \varphi_1 + \theta_2 \varphi_2 + \theta_2 \theta_1 v$ is a superfield.

\item $u(x,t)$ and $v(x,t)$ are bosonic fields.

\item $\varphi_1(x,t)$ and $\varphi_2(x,t)$ are fermionic fields. 

\item $A = i D_1 D_2 - \gamma$.

\item $M = AU$.
\end{itemize}
Equations (\ref{CH}) and (\ref{HS}) can then be combined into the single equation
\begin{equation}\label{CHHS}
               m_t = -2u_x m - um_x, \qquad x\in\R,\;t>0.
\end{equation}
Although there exist several $N = 2$ supersymmetric extensions of (\ref{CHHS}), some generalizations have the particular property that their bosonic sectors are equivalent to the most popular two-component generalizations of (\ref{CH}) and (\ref{HS}) given by (see e.g. \cite{C-L-Z, E-L-Y, Ivanov})
\begin{equation}\label{2CHHS}
 \begin{cases}
 	m_t + 2u_x m + um_x + \sigma \rho \rho_x = 0, \qquad \sigma = \pm 1,
		\\
	\rho_t + (u \rho)_x = 0.
	\end{cases}
\end{equation}
The $N =2$ supersymmetric generalization of equation (\ref{CHHS}) that we will consider in this paper shares this property and is given by
\begin{equation}\label{superCHHS}
  M_t = -(M U )_x + \frac{1}{2}\left[(D_1M) (D_1U) + (D_2M) (D_2U)\right].
\end{equation}  
Defining $\rho$ by 
\begin{equation}\label{rhodef}
\rho =\begin{cases}
   v + i\gamma u 	&\text{ if }  \sigma = 1,
	\\
  i v - \gamma u	&\text{ if } \sigma = -1,
\end{cases}
\end{equation}
the bosonic sector of (\ref{superCHHS}) is exactly the two-component equation (\ref{2CHHS}).

If $u$ and $\rho$ are allowed to be complex-valued functions, the two versions of (\ref{2CHHS}) corresponding to $\sigma = 1$ and $\sigma = -1$ are equivalent, because the substitution $\rho \to i \rho$ converts one into the other. However, in the usual context of real-valued fields, the equations are distinct. The discussion in \cite{P2006} focused attention on the two-component generalization with $\sigma = -1$. 
The observation that the $N=2$ supersymmetric Camassa-Holm equation  
arises as an Euler equation was already made in~\cite{AGZ}.

\section{Geodesic flow} \nequation
As noted above, equations (\ref{CH}) and (\ref{HS}) allow geometric interpretations as equations for geodesic flow related to the group of diffeomorphisms of the circle $S^1$ endowed with a right-invariant metric.\footnote{In this section we consider all equations within the spatially periodic setting---although formally the same arguments apply to the case on the line, further technical complications arise due to the need of imposing boundary conditions at infinity cf. \cite{C2}.}
More precisely, using the right-invariance of the metric, the full geodesic equations can be reduced by symmetry to a so-called Euler equation in the Lie algebra of vector fields on the circle together with a reconstruction equation \cite{M-R} (see also \cite{Kolev, Kolev2}). This is how equations (\ref{CH}) and (\ref{HS}) arise as Euler equations related to the algebra $\text{Vect}(S^1)$.

In this section we describe how equation (\ref{superCHHS}) similarly arises as an Euler equation related to the superconformal algebra $K(S^{1|2})$ of contact vector fields on the $1|2$-dimensional supercircle $S^{1|2}$. Since $K(S^{1|2})$ is related to the group of superdiffeomorphisms of $S^{1|2}$, this leads (at least formally) to a geometric interpretation of equation (\ref{superCHHS}) as an equation for geodesic flow.

\subsection{The superconformal algebra $K(S^{1|2})$}
The Lie superalgebra $K(S^{1|2})$ is defined as follows cf. \cite{D-M, G-R}. The supercircle $S^{1|2}$ admits local coordinates $x, \theta_1, \theta_2$, where $x$ is a local coordinate on $S^1$ and $\theta_1, \theta_2$ are odd coordinates. 
Let $\text{Vect}(S^{1|2})$ denote the set of vector fields on $S^{1|2}$. An element $X \in \text{Vect}(S^{1|2})$ can be written as
$$X = f(x,\theta_1, \theta_2) \frac{\partial}{\partial x} + f^1(x,\theta_1, \theta_2) \frac{\partial}{\partial \theta_1} + f^2(x,\theta_1, \theta_2) \frac{\partial}{\partial \theta_2},$$
where $f, f^1, f^2$ are functions on $S^{1|2}$. Let 
$$\alpha = dx + \theta_1 d\theta_1 + \theta_2 d\theta_2,$$
be the contact form on $S^{1|2}$. The superconformal algebra $K(S^{1|2})$ consists of all contact vector fields on $S^{1|2}$, i.e.
$$K(S^{1|2}) = \left\{X \in \text{Vect}(S^{1|2}) \bigl| \, L_X \alpha = f_X \alpha \text{    for some function $f_X$ on $S^{1|2}$}\right\},$$
where $L_X$ denotes the Lie derivative in the direction of $X$.

A convenient description of $K(S^{1|2})$ is obtained by viewing its elements as Hamiltonian vector fields corresponding to functions on $S^{1|2}$. Indeed, define the Hamiltonian vector field $X_f$ associated with a function $f(x, \theta_1, \theta_2)$ by
$$X_f = (-1)^{|f| +1} \left(\frac{\partial f}{\partial \theta_1}\frac{\partial }{\partial \theta_1} + \frac{\partial f}{\partial \theta_2}\frac{\partial }{\partial \theta_2}\right),$$
where $|f|$ denotes the parity of $f$. The Euler vector field is defined by
$$E = \theta_1 \frac{\partial }{\partial \theta_1} + \theta_2 \frac{\partial }{\partial \theta_2},$$
and, for each function $f$ on $S^{1|2}$, we let $D(f) = 2f - E(f)$. Then the map 
$$f \mapsto K_f := D(f) \frac{\partial }{\partial x} - X_f + \frac{\partial f}{\partial x} E,$$
satisfies $[K_f, K_g] = K_{\{f, g\}}$ where
\begin{equation}\label{fgbracket}
  \{f, g\} = D(f) \frac{\partial g}{\partial x} - \frac{\partial f}{\partial x}D(g) + (-1)^{|f|}\left(\frac{\partial f}{\partial \theta_1}\frac{\partial g}{\partial \theta_1} + \frac{\partial f}{\partial \theta_2}\frac{\partial g}{\partial \theta_2}\right).
\end{equation}
Thus, the map $f\to K_f$ is a homomorphism from the Lie superalgebra of 
functions $f$ on $S^{1|2}$ endowed with the bracket (\ref{fgbracket})
to $K(S^{1|2})$.

\subsection{Euler equation on $K(S^{1|2})$}
For two even functions $U$ and $V$ on $S^{1|2}$ (we usually refer to $U$ and $V$ as `superfields'), we define a Lie bracket $[ \cdot, \cdot ]$ by
\begin{equation}\label{liebracket}
  [U, V] = UV_x - U_x V + \frac{1}{2}\left[(D_1U)(D_1V) + (D_2U)(D_2V)\right].
\end{equation}
It is easily verified that $[U, V] = \frac{1}{2}\{U, V\}$, where $\{\cdot, \cdot\}$ is the bracket in (\ref{fgbracket}).
The Euler equation with respect to a metric $\langle \cdot, \cdot \rangle$ is given by \cite{A}
\begin{equation}\label{UtBUU}  
  U_t = B(U, U),
\end{equation}
where the bilinear map $B(U,V)$ is defined by the relation
$$\langle B(U, V), W \rangle = \langle U, [V, W] \rangle,$$
for any three even superfields $U, V, W$.
Recall that $A = i D_1 D_2 - \gamma$. Letting
\begin{equation}\label{Ametric}
  \langle U, V \rangle = -i \int dx d\theta_1 d\theta_2 U AV,
\end{equation}  
a computation shows that
\begin{equation}\label{BUV}  
  B(U,U) = A^{-1}\left[-\left(MU\right)_x + \frac{1}{2}\left[(D_1M)(D_1U) + (D_2M)(D_2U) \right]\right].
\end{equation}
It follows from (\ref{UtBUU}) and (\ref{BUV}) that (\ref{superCHHS}) is the Euler equation corresponding to $\langle \cdot, \cdot \rangle$ given by (\ref{Ametric}).

\begin{remark}\upshape
We make the following observations:

1. The action of the inverse $A^{-1}$ of the operator $A = i D_1 D_2 - \gamma$ in equation (\ref{BUV}) is well-defined. Indeed, the map $A$ can be expressed as
\begin{equation}\label{Acomponents}
A: \begin{pmatrix} u \\ \varphi_1 \\ \varphi_2 \\ v \end{pmatrix} \mapsto \begin{pmatrix} iv - \gamma u \\ i \varphi_{2x} - \gamma \varphi_1 \\ -i\varphi_{1x} - \gamma \varphi_2 
\\ -iu_{xx} - \gamma v \end{pmatrix},
\end{equation}
where we identify a superfield with the column vector made up of its four component fields (e.g. $U = u  + \theta_1 \varphi_1 + \theta_2 \varphi_2 + \theta_2 \theta_1 v$ is identified with the column vector $(u, \varphi_1, \varphi_2,v)^T$). Let $C^\infty(S^1; \C)$ denote the space of smooth periodic complex-valued functions. Since the operator $1 - \partial_x^2$ is an isomorphism from $C^\infty(S^1; \C)$ to itself, we deduce from (\ref{Acomponents}) that the operation of $A^{-1}$ on smooth periodic complex-valued superfields is well-defined when $\gamma = 1$. 

Consider now the case of $\gamma =0$. In this case it follows from (\ref{Acomponents}) that $A^{-1}$ involves the inverses of the operators $\partial_x^2$ and $\partial_x$. In order to make sense of these inverses we restrict the domain of $A$ to the set 
$$\mathcal{E} = \{U = u  + \theta_1 \varphi_1 + \theta_2 \varphi_2 + \theta_2 \theta_1 v \text{ is smooth and periodic} | u(0) = \varphi_1(0) = \varphi_2(0) = 0\}.$$
The operator $A = i D_1 D_2$ maps this restricted domain $\mathcal{E}$ bijectively onto the set 
\begin{align*}
\mathcal{F} = \biggl\{M = i(n  + \theta_1 \psi_1 + \theta_2 \psi_2 + \theta_2 \theta_1 m) &\text{ is smooth and periodic} \biggl| 
	\\
&\int_{S^1} m dx = \int_{S^1} \psi_1 dx =\int_{S^1} \psi_2 dx =0\biggr\}.
\end{align*}
Hence the inverse $A^{-1}$ is a well-defined map $\mathcal{F} \to \mathcal{E}$. Writing $M = i(n  + \theta_1 \psi_1 + \theta_2 \psi_2 + \theta_2 \theta_1 m)$, $A^{-1}M$ is given explicitly by 
$$A^{-1}M  = \begin{pmatrix}
 -\int_0^x\int_0^y m(z)dzdy +  x \int_{S^1}\int_0^y m(z)dz dy 
	 \\
 -\int_0^x \psi_2(y) dy
 	\\
\int_0^x \psi_1(y) dy 
 	\\
 n(x)  
 \end{pmatrix}.$$
Note that the expression 
$$-\left(MU\right)_x + \frac{1}{2}\left[(D_1M)(D_1U) + (D_2M)(D_2U) \right]
= -i \left[(D_1D_2U)U + \frac{1}{2}(D_1U)(D_2U)\right]_x$$ 
acted on by $A^{-1}$ in equation (\ref{BUV}) belongs to $\mathcal{F}$ for any even superfield $U$. Hence the operation of $A^{-1}$ in equation (\ref{BUV}) is well-defined also when $\gamma = 0$. 
The restriction of the domain of $A$ to $\mathcal{E}$ is related to the fact that the two-component equation (\ref{2CHHS}) with $\gamma = 0$ is invariant under the symmetry
$$u(x,t) \to u(x - c(t),t) + c'(t), \qquad \rho(x,t) \to \rho(x - c(t),t),$$
for any sufficiently regular function $c(t)$. Hence, by enforcing the condition $u(0) = 0$ we remove obvious non-uniqueness of the solutions to the equation.

2. If we restrict attention to the bosonic sector and let
$$U = u_1  + \theta_2 \theta_1 v_1, \qquad V = u_2  + \theta_2 \theta_1 v_2,$$
the Lie bracket (\ref{liebracket}) induces on two pairs of functions $(u_1,v_1)$ and $(u_2,v_2)$ the bracket
\begin{equation}\label{liebracketbosonic}
   \left[(u_1, v_1), (u_2, v_2)\right] = \left(u_1 u_{2x} - u_{1x} u_2, u_1v_{2x} - u_2v_{1x}\right).
\end{equation}
We recognize (\ref{liebracketbosonic}) as the commutation relation for the semidirect product Lie algebra $\hbox{Vect}(S^1) \ltimes C^\infty(S^1)$, where $\hbox{Vect}(S^1)$ denotes the space of smooth vector fields on $S^1$, see \cite{G-O}. 

3. The restriction of the metric (\ref{Ametric}) to the bosonic sector induces on pair of functions $(u,v)$ the bilinear form 
\begin{equation}\label{metricbosonic}
  \langle (u_1, v_1), (u_2, v_2)\rangle = \int \left(i \gamma(u_1 v_2 + u_2 v_1) + u_{1x} u_{2x} + v_1 v_2\right) dx.
\end{equation}
Changing variables from $(u_j, v_j)$ to $(u_j, \rho_j)$, $j = 1, 2$, according to (\ref{rhodef}), equation (\ref{metricbosonic}) becomes
\begin{equation}\label{metricbosonic2}
\langle (u_1, \rho_1), (u_2, \rho_2)\rangle = \int \left(\gamma u_1 u_2 + u_{1x} u_{2x} + \sigma \rho_1 \rho_2\right) dx.
\end{equation}
Hence, as expected, when $\rho_1 = \rho_2 = 0$ the metric reduces to the $H^1$ metric (\ref{H1metric}) in the case of CH, while it reduces to the $\dot{H}^1$ metric (\ref{H1dotmetric}) in the case of HS.

4. Our discussion freely used complex-valued expressions and took place only at the Lie algebra level. The extent to which there actually exists a corresponding geodesic flow on the superdiffeomorphism group of the supercircle $S^{1|2}$ has to be further investigated. 
A first step towards developing such a geometric picture involves  
dealing with the presence of imaginary numbers in the definition of the  
metric (\ref{Ametric}) when $\gamma = 1$. Although these imaginary factors disappear in  
the bosonic sector when changing variables to $(u, \rho)$ (see  
(\ref{metricbosonic2})), they are still present in the fermionic sector. 
It seems unavoidable to encounter complex-valued expressions at one point or  
another of the present construction if one insists on the system being  
an extension of equation (\ref{CH}) (in the references \cite{AGZ} and  
\cite{P2006} imaginary factors appear when the coefficients are chosen  
in such a way that the system is an extension of (\ref{CH})).
\end{remark}

\section{Bi-Hamiltonian formulation}\label{bihamsec} \nequation
Equation (\ref{superCHHS}) admits the bi-Hamiltonian structure\footnote{By definition the variational derivative $\delta H/\delta M$ of a functional $H[M]$ is required to satisfy
$$\frac{d}{d\epsilon} H[M + \epsilon \delta M]\biggr|_{\epsilon =0} = \int dx d\theta_1 d\theta_2 \frac{\delta H}{\delta M} \delta M,$$
for any smooth variation $\delta M$ of $M$. }
\begin{equation}\label{biham}
  M_t =  J_1 \frac{\delta H_1}{\delta M} = J_2 \frac{\delta H_2}{\delta M},
\end{equation}  
where the Hamiltonian operators $J_1$ and $J_2$ are defined by
\begin{align}
J_1 = i\left[-\partial_x M - M \partial_x + \frac{1}{2}\left(D_1 M D_1 + D_2 M D_2\right)\right],
	\qquad
J_2 = -i\partial_x A,
\end{align}
and the Hamiltonian functionals $H_1$ and $H_2$ are defined by
\begin{equation}
H_1 = -\frac{i}{2} \int dx d\theta_1 d\theta_2 M U, \qquad  H_2 =
-\frac{i}{4} \int dx d\theta_1 d\theta_2 \left(M U^2 - \frac{\gamma}{3} U^3\right).
\end{equation}
The bi-Hamiltonian formulation (\ref{biham}) is a particular case of a construction in \cite{P2006}, where it was verified that the operators $J_1$ and $J_2$ are compatible. The first few conservation laws in the hierarchy generated by $J_1$ and $J_2$ are presented in Figure \ref{superhierarchyfig}.

 \begin{figure}
\begin{center}
$$ \xymatrix{
  &  \ar[dl]_{J_2}   \frac{\delta}{\delta M} \left[-\frac{i}{4} \int dx d\theta_1 d\theta_2 \left(M U^2 - \frac{\gamma}{3} U^3\right) \right]= \frac{\delta H_2}{\delta M} 	
  		\\
\text{Equation (\ref{superCHHS})}
 & & 
 		\\  
    &   \ar[ul]_{J_1} \ar[dl]_{J_2}   \frac{\delta}{\delta M} \left[-\frac{i}{2} \int dx d\theta_1 d\theta_2 M U \right] = \frac{\delta H_1}{\delta M}
    \\
M_t = -M_x &  
 		\\
  &   \ar[ul]_{J_1} \ar[dl]_{J_2}   \frac{\delta}{\delta M}\left[-i \int dx d\theta_1 d\theta_2 M \right] = \frac{\delta H_0}{\delta M}
  		\\
	M_t = 0  &  	
		 }$$
     \begin{figuretext}\label{superhierarchyfig}
       Recursion scheme for the operators $J_1$ and $J_2$ associated with the supersymmetric equation (\ref{superCHHS}).
     \end{figuretext}
     \end{center}
\end{figure}

Restricting attention to the purely bosonic sector of the bi-Hamiltonian structure of (\ref{superCHHS}), we recover bi-Hamiltonian formulations for the two-component generalizations of (\ref{CH}) and (\ref{HS}). More precisely, we find that equation (\ref{2CHHS}) can be put in the bi-Hamiltonian form\footnote{The gradient of a functional $F[m,\rho]$ is defined by
$$\text{grad}\, F = \begin{pmatrix} \frac{\delta F}{\delta m} \\ \frac{\delta F}{\delta \rho} \end{pmatrix},$$
provided that there exist functions $\frac{\delta F}{\delta m}$ and $\frac{\delta F}{\delta \rho}$ such that
$$\frac{d}{d\epsilon} F[m + \epsilon \delta m, \rho + \epsilon \delta \rho]\biggr|_{\epsilon =0} = \int \biggl(\frac{\delta F}{\delta m} \delta m + \frac{\delta F}{\delta \rho} \delta \rho \biggr)dx,$$
for any smooth variations $\delta m$ and $\delta \rho$.}
\begin{equation}\label{2biham}
  \begin{pmatrix} m \\ \rho \end{pmatrix}_t =  K_1 \text{grad}\, G_1 = K_2 \text{grad}\, G_2,
\end{equation}  
where the Hamiltonian operators are defined by
\begin{align}
K_1 =  \begin{pmatrix} -m\partial_x - \partial_x m	&	-\rho \partial
	\\
	-\partial \rho	&	0 \end{pmatrix},
	\qquad
K_2  = 
 \begin{pmatrix} -\partial \Lambda  &0
 \\ 0 & -\sigma\partial \end{pmatrix}, 
\end{align}
and the Hamiltonians $G_1$ and $G_2$ are given by
$$G_1 = \frac{1}{2} \int (um + \sigma \rho^2) dx, \qquad G_2 = \frac{1}{2}\int (\sigma u\rho^2 + \gamma u^3 + u u_x^2) dx.$$
For $j = 1,2$, $G_j$ are the restrictions to the purely bosonic sector
of the functionals $H_j$.

 \begin{figure}
\begin{center}
$$ \xymatrix{
  &  \ar[dl]_{K_2}   \text{grad}\, \frac{1}{2} \int \left(\sigma u\rho^2 + \gamma u^3 + u u_x^2 \right) dx = \text{grad}\, G_2	
  		\\
{\begin{pmatrix} m \\ \rho \end{pmatrix}_t
=  \begin{pmatrix} -2u_x m - um_x - \sigma \rho \rho_x \\
-(u\rho)_x \end{pmatrix}} & & 
 		\\  
    &   \ar[ul]_{K_1} \ar[dl]_{K_2}   \text{grad}\, \frac{1}{2} \int \left(um + \sigma \rho^2\right) dx = \text{grad}\, G_1\\
{\begin{pmatrix} m \\ \rho \end{pmatrix}_t
= - \begin{pmatrix} m_x  \\
 \rho_x \end{pmatrix}}  &  
 		\\
  &   \ar[ul]_{K_1} \ar[dl]_{K_2}   \text{grad}\, \int m dx = \text{grad}\, G_0
  		\\
{\begin{pmatrix} m \\ \rho \end{pmatrix}_t
= \begin{pmatrix} 0 \\
 0 \end{pmatrix}}   & 
		 \\
 &   \ar[ul]_{K_1}  \ar[dl]_{K_2}    \text{grad}\, \int \rho dx = \text{grad}\, G_{-1}	
		 \\
{\begin{pmatrix} m \\ \rho \end{pmatrix}_t
= \begin{pmatrix} 0 \\
 0 \end{pmatrix}}   &   
 		 \\
 &   \ar[ul]_{K_1}   \ar[dl]_{K_2}   \text{grad}\, \int \frac{m}{\rho} dx = \text{grad}\, G_{-2}
 		\\
{\begin{pmatrix} m \\ \rho \end{pmatrix}_t
= \begin{pmatrix} 
 \frac{\gamma \rho_x}{\rho^2} + \left(\frac{1}{\rho}\right)_{xxx}
 \\
\sigma\left(\frac{m}{\rho^2}\right)_x
 \end{pmatrix}}   &   
 		 \\
 &   \ar[ul]_{K_1}     \text{grad}\, \int \left( \frac{-\sigma m^2}{2 \rho^3} - \frac{\gamma}{2\rho}  - \frac{\rho_{xx}}{4 \rho^2} \right) dx
  = \text{grad}\, G_{-3}
		 }$$
     \begin{figuretext}\label{2hierarchyfig}
       Recursion scheme for two-component generalization (\ref{2CHHS}) of the CH (corresponding to $m = u-u_{xx}$) and the HS (corresponding to $m = -u_{xx}$) equations.
     \end{figuretext}
     \end{center}
\end{figure}

The hierarchy of conserved quantities $G_n$ for the two-component system (\ref{2CHHS}) can be obtained from the recursive relations
$$K_1 \text{grad}\, G_n = K_2 \text{grad}\, G_{n+1}, \qquad n \in \Z.$$
The associated commuting Hamiltonian flows are given by
$$\begin{pmatrix} m \\ \rho \end{pmatrix}_t = K_1 \text{grad}\, G_{n} = K_2 \text{grad}\, G_{n+1}, \qquad n \in \Z.$$
Since $K_2^{-1}$ is a nonlocal operator the expressions for members $G_n$ with $n \geq 3$ are nonlocal when written as functionals of $(u, \rho)$.
On the other hand, since the operator $K_1$ can be explicitly inverted as
$$K_1^{-1} =  \begin{pmatrix} 0	&	-\frac{1}{\rho} \partial_x^{-1}
	\\
	-\partial_x^{-1} \frac{1}{\rho}	&	\partial_x^{-1} \frac{1}{\rho}(m\partial_x + \partial_x m)\frac{1}{\rho} \partial_x^{-1} \end{pmatrix},$$
it is possible to implement on a computer a recursive algorithm for finding the conservation laws $G_n$ for $n \leq 2$. The first few of these conserved quantities and their associated Hamiltonian flows are presented in Figure \ref{2hierarchyfig}. Note that $G_0$ is a Casimir for the positive hierarchy, $G_{-2}$ is a Casimir for the negative hierarchy, and $G_{-1}$ is a Casimir for both the positive and negative hierarchies.

This discussion illustrates that even though many properties of the two-component system (\ref{2CHHS}) carry over to its supersymmetric extension (\ref{superCHHS}) (such as the property of being a geodesic equation as shown above, and the Lax pair formulation as shown below), the construction of a negative hierarchy does not appear to generalize. Certainly, since the conservation laws $G_{-1}$, $G_{-2}$, ..., contain negative powers of $\rho$, these functionals cannot be the restriction to the bosonic sector of functionals $H_{-1}$, $H_{-2}$, ..., which involve only polynomials of the fields $U$ and $M$ and derivatives of these fields.

\section{Lax pair} \nequation
Equation (\ref{superCHHS}) is the condition of compatibility of the linear system
\begin{equation}\label{superlax}
\begin{cases}
  i D_1 D_2 G = \left(\frac{1}{2\lambda}M + \frac{\gamma}{2}\right)G,
  	\\
  G_t = \frac{1}{2} U_x G - \frac{1}{2}\left[(D_1U)(D_1G) + (D_2U)(D_2G)\right] - (\lambda + U)G_x,
\end{cases}
\end{equation}
where the even superfield $G$ serves as an eigenfunction and $\lambda \in \C$ is a spectral parameter.
The Lax pair (\ref{superlax}) can be derived from the bi-Hamiltonian formulation presented in Section \ref{bihamsec} as follows. In view of the general theory of recursion operators \cite{F1987}, we expect
\begin{equation}\label{J1FJ2F}
  J_1 F = \lambda J_2 F,
\end{equation}
to be the $x$-part of a Lax pair for (\ref{superCHHS}). This Lax pair involves the even superfield $F$, which serves as a `squared eigenfunction' cf. \cite{F-A}.
Letting $F = G^2$, a computation shows that equation (\ref{J1FJ2F}) holds whenever $G$ satisfies the $x$-part of (\ref{superlax}). 
In order to find the corresponding $t$-part, we make the Ansatz
\begin{equation}\label{tpartAnsatz}
  G_t = A_1 G + B_1 D_1G + B_2 D_2G + A_2G_x,
\end{equation}
where $A_1, A_2$ ($B_1, B_2$) are even (odd) superfields.
Long but straightforward computations show that the compatibility of (\ref{tpartAnsatz}) with the $x$-part of (\ref{superlax}) is equivalent to the following four equations:
\begin{align} \label{comp1}
M_t 
=& 2\lambda i D_1D_2A_1  + (D_1B_1 + D_2 B_2)(M+ \lambda \gamma)
	\\ \nonumber
&+ (D_1A_2 - B_1)D_1M + (D_2A_2 - B_2)D_2M + A_2 M_x,
	\\ \label{comp2}
\left(\frac{1}{2\lambda} M + \frac{\gamma}{2}\right) B_1 =& iD_1D_2B_1 - i D_2A_1  + (D_1A_2 - B_1)\left(\frac{1}{2\lambda} M + \frac{\gamma}{2}\right),
		\\\label{comp3}
\left(\frac{1}{2\lambda} M + \frac{\gamma}{2}\right) B_2 =& iD_1D_2B_2 + i D_1A_1 + (D_2A_2 - B_2)\left(\frac{1}{2\lambda} M + \frac{\gamma}{2}\right),
	\\\label{comp4}
0 =& iD_1D_2A_2 - i D_1B_2 + i D_2B_1.
\end{align}
Equations (\ref{comp1})-(\ref{comp4}) ascertain the equality of the coefficients of $G, D_1G, D_2G, G_x$, respectively, on the left- and right-hand sides of the compatibility equation $i D_1 D_2 (G_t) = (i D_1 D_2 G)_t$.
Letting
$$A_1 = \frac{1}{2}U_x, \qquad B_1 = -\frac{1}{2}D_1U, 
\qquad B_2 = -\frac{1}{2}D_2U, \qquad A_2 = -(\lambda + U),$$
equations (\ref{comp2})-(\ref{comp4}) are identically satisfied, while equation (\ref{comp1}) reduces to the supersymmetric system (\ref{superCHHS}).

\section{Explicit solutions: First deconstruction}\label{explicitfirstsec} \nequation
In order to obtain explicit solutions of the supersymmetric system (\ref{superCHHS}), we consider the simplest case of fields taking values in the Grassmann algebra consisting of two elements $1$ and $\tau$ satisfying the following relations:
\begin{equation}\label{taurelations}
  1^2 = 1, \qquad \tau \cdot 1 = \tau = 1 \cdot \tau, \qquad \tau^2 = 0.
\end{equation}  
In the case of this so-called first deconstruction the fields can be written as
$$u = u \cdot 1, \qquad \rho = \rho \cdot 1, \qquad \varphi_1 = \tau \cdot f_1, \qquad \varphi_2 = \tau \cdot f_2,$$
where $f_1(x,t)$ and $f_2(x,t)$ are ordinary real-valued functions. The component equations of (\ref{superCHHS}) consist of the bosonic two-component system (\ref{2CHHS}) together with the following equations for the fermionic fields:
\begin{align}\label{travsystems1}
 & \begin{cases}
    \left(f_1 \rho +2 f_{2t} + 2 u f_{2x} + f_2 u_x \right)_x
     +  i  \gamma  \left(2 f_{1t} + 3 u f_{1x} + 3 f_1 u_x \right)
     = 0,
   	\\
	\left(- f_2 \rho +2 f_{1t} + 2 u f_{1x} + f_1 u_x \right)_x 
	   - i \gamma  \left(2 f_{2t} + 3 u f_{2x} + 3 f_2 u_x \right) = 0,
  \end{cases}
& \sigma = 1,
  \\ \label{travsystems2}
&  \begin{cases}
   \left(-if_1 \rho + 2 f_{2t} + 2u f_{2x} + f_2 u_x \right)_x + i \gamma  \left(2 f_{1t}+3 u f_{1x}+3 f_1 u_x\right) = 0,
	\\   
 \left( i f_2 \rho + 2 f_{1t} + 2u f_{1x} + f_1 u_x \right)_x - i   \gamma  \left(2 f_{2t}+3 u f_{2x}+3f_2 u_x\right) = 0,
  \end{cases}
& \sigma = -1.
\end{align}

We seek traveling-wave solutions for which $u$ and $\rho$ are of the form 
\begin{equation}\label{urhotrav}
  u = u(y), \qquad \rho = \rho(y), \qquad y = x - ct,
\end{equation}
where $c \in \R$ denotes the velocity of the wave. Substituting (\ref{urhotrav}) into (\ref{2CHHS}), we find
\begin{equation}\label{2CHHStrav}
 \begin{cases}
 	-c m_y + 2u_y m + um_y + \sigma \rho \rho_y = 0, \qquad \sigma = \pm 1,
		\\
	-c \rho_y + (u \rho)_y = 0.
	\end{cases}
\end{equation}
The second of these equations yields, for some constant of integration $a$,
\begin{equation}\label{travrho}  
  \rho(y) = \frac{a}{c - u(y)}.
\end{equation}
At this point it is convenient to restrict attention to the case of either CH or HS. We choose to consider solutions of the HS system.

\subsection{Traveling waves for the supersymmetric HS equation}
We first derive solutions for the bosonic fields before considering the extension to the fermionic sector. Since the bosonic sector coincides with equation (\ref{2CHHS}), this analysis yields as a by-product a classification of the bounded traveling-wave solutions of the two-component HS equation.

\subsubsection{Bosonic fields}
Substituting expression (\ref{travrho}) for $\rho$ into the first equation of (\ref{2CHHStrav}) and integrating the resulting equation with respect to $y$, we find
\begin{equation}\label{cuyytrav}  
  c u_{yy} - \frac{1}{2}u_y^2 - uu_{yy} + \frac{\sigma a^2}{2(c - u)^2} = \frac{b}{2},
\end{equation}
where $b$ is an arbitrary constant of integration.
We multiply this equation by $2u_y$ and integrate the resulting equation with respect to $y$. After simplification we arrive at the ordinary differential equation
\begin{equation}\label{HStravODE}  
  u_{y}^2 = \frac{(b u + d)(c - u) - \sigma a^2}{(c - u)^2}, 
\end{equation}
where $d$ is yet another integration constant. 
Introducing $\umin$ and $\umax$ so that 
\begin{equation}\label{uminumaxdef}  
  (b u + d)(c - u) - \sigma a^2 = b(\umax - u)(u - \umin),
\end{equation}
an analysis of (\ref{HStravODE}) reveals that bounded traveling waves arise when $\umin, \umax, c, b$ satisfy\footnote{See \cite{L3} for more details of a similar analysis in the case of equation (\ref{CH}).}  
\begin{equation}\label{bzZcineq}
\text{$b > 0$ and either $\umin < \umax < c$ or $c < \umin < \umax$.}
\end{equation}
If (\ref{bzZcineq}) is fulfilled, then there exists a smooth periodic solution $u(y)$ of equation (\ref{HStravODE}) such that
$$\min_{y \in \R} u(y) = \umin, \qquad \max_{y \in \R} u(y) = \umax.$$
We obtain a traveling wave of the two-component HS equation by constructing $\rho(y)$ from $u(y)$ according to (\ref{travrho}). 
It can be checked by considering all possible distributions of the parameters $b, \umin, \umax, c$ that there are no other bounded continuous traveling-wave solutions of the two-component HS equation.\footnote{If $b > 0$ and $\umin < c < \umax$ there exist periodic solutions $u(y)$ of (\ref{HStravODE}) with cusps, where we say that a continuous function $u(y)$ has a cusp at $y = y_0$ if $u$ is smooth on either side of $y_0$ but $u_y \to \pm \infty$ as $y \to y_0$. However, the existence of a cusp of $u$ at $y = y_0$ implies that $u(y_0) = c$ so that the corresponding function $\rho(y)$ given by (\ref{travrho}) is unbounded near $y_0$. This argument explains why these solutions are excluded from the list of bounded traveling waves. Similarly, the cusped solutions of (\ref{HStravODE}) which exist when $b < 0$ and either $z < Z < c$ or $c < z < Z$ do not give rise to bounded traveling waves.}

The relations
$$b(\umin + \umax) = bc - d, \qquad -b\umin\umax = dc - \sigma a^2,$$
obtained from (\ref{uminumaxdef}), show that
\begin{equation}\label{bcsigmaa}  
  b(c- \umin)(c-\umax) = \sigma a^2.
\end{equation}
There are two cases to consider: $a = 0$ and $a \neq 0$. If $a = 0$, then either $b = 0$, $\umin=c$ or $\umax=c$, so that (\ref{bzZcineq}) does not hold. Hence no traveling waves exist in this case. If $a \neq 0$, then $a^2$ is a strictly positive number and relation (\ref{bcsigmaa}) requires that
$$b(c- \umin)(c-\umax) \gtrless 0 \quad \hbox{for} \quad \sigma = \pm 1.$$
The left-hand side of this equation is positive whenever (\ref{bzZcineq}) holds. This shows that no traveling waves exist when $\sigma = -1$. On the other hand, if $\sigma =1$ any combination of $b, \umin, \umax, c \in \R$ satisfying (\ref{bzZcineq}) corresponds to exactly two unique smooth periodic traveling-wave solutions. 
There are {\it two} solutions for each admissible combination of $b, \umin, \umax, c$ because these parameters determine $a$, and hence $\rho$, only up to sign. In general, equation (\ref{2CHHS}) is invariant under the symmetry
$$u \to u, \qquad \rho \to -\rho.$$

Let us mention that equation (\ref{HS}) admits no bounded traveling waves. This can be seen from the above analysis because $\rho = 0$ if and only if $a = 0$ and we have seen that no traveling waves exist in this case. Hence, in this regard the two-component generalization with $\sigma =1$ presents a qualitatively richer structure than its one-component analogue.

\subsubsection{Fermionic fields}
We now consider the fermionic fields corresponding to the periodic traveling waves found in the previous subsection. Hence let $\sigma = 1$ and let $(u, \rho)$ be a smooth periodic traveling-wave solution of the two-component HS equation (\ref{2CHHS}). The relevant fermionic equations obtained from (\ref{travsystems1}) and (\ref{travrho}) are
\begin{equation}\label{HStravfermionic}  
 \begin{cases}
    \frac{a f_1}{c - u} + 2 f_{2t} + 2u f_{2x} + f_2 u_x  = B,
	\\   
  \frac{-af_2}{c - u} + 2 f_{1t} + 2u f_{1x} + f_1 u_x  = C,
  \end{cases}
\end{equation}
where $B$ and $C$ are arbitrary constants of integration. One way to obtain explicit (complex-valued) solutions of (\ref{HStravfermionic}) is to seek solutions of the form $B = C = 0$ and $f_1 = if_2 =: f(y)$, where, as above, $y = x - ct$. In this case the system (\ref{HStravfermionic}) reduces to the single equation
\begin{equation}\label{HStravfermionicsimple}  
  \frac{i af}{c - u} - 2(c - u) f_y + f u_y = 0,
\end{equation}
This equation can be solved explicitly for $f$. Equations (\ref{HStravODE}), (\ref{uminumaxdef}), and (\ref{HStravfermionicsimple}) yield
\begin{equation*}
 \left( \frac{ia + \text{sgn}((c-u)u_y)\sqrt{b(\umax-u)(u-\umin)}}{c - u}\right) f - 2(c - u) f_y = 0.
\end{equation*}
Using that $du = \text{sgn}((c-u)u_y)\frac{\sqrt{b(\umax-u)(u-\umin)}}{c - u} dy$, we can write the solution to this equation as 
\begin{equation}\label{logfint}
  \log f = \int_{u_0}^u \frac{ia + \text{sgn}((c-u)u_y)\sqrt{b(\umax-u)(u-\umin)}}{2(c - u)} \frac{du}{\text{sgn}((c-u)u_y)\sqrt{b(\umax-u)(u-\umin)}}.
\end{equation}  
We find $f$ by computing the integral in (\ref{logfint}) and using the relation $a = \text{sgn}(a) \sqrt{b(c-\umin)(c-\umax)}$ to eliminate $a$. Letting $f_0$ denote the value of $f$ at a point where $u = \umin$, we obtain
\begin{equation}\label{fyf0sqrt}
f(y) = f_0 \sqrt{\frac{c-\umin}{c-u(y)}} e^{\frac{i}{2}\text{sgn}(a (c-u)u_y)\arctan\left(\frac{2 \sqrt{(c-\umin)(c-\umax)(\umax-u(y))(u(y)-\umin)}}{-2 \umin \umax+c (\umin+\umax-2 u(y))+(\umin+\umax) u(y)}\right)}.
\end{equation}
Note that the identity 
$$e^{\pm \frac{i}{2} \arctan t} = \left(\frac{1 + i t}{1 - it}\right)^{\pm 1/4},$$
implies that apart from the presence of fractional powers the right-hand side of (\ref{fyf0sqrt}) is a rational function of $u$.

\subsubsection{Summarized result}
\begin{figure}
\begin{center}
    \includegraphics[width=.90\textwidth]{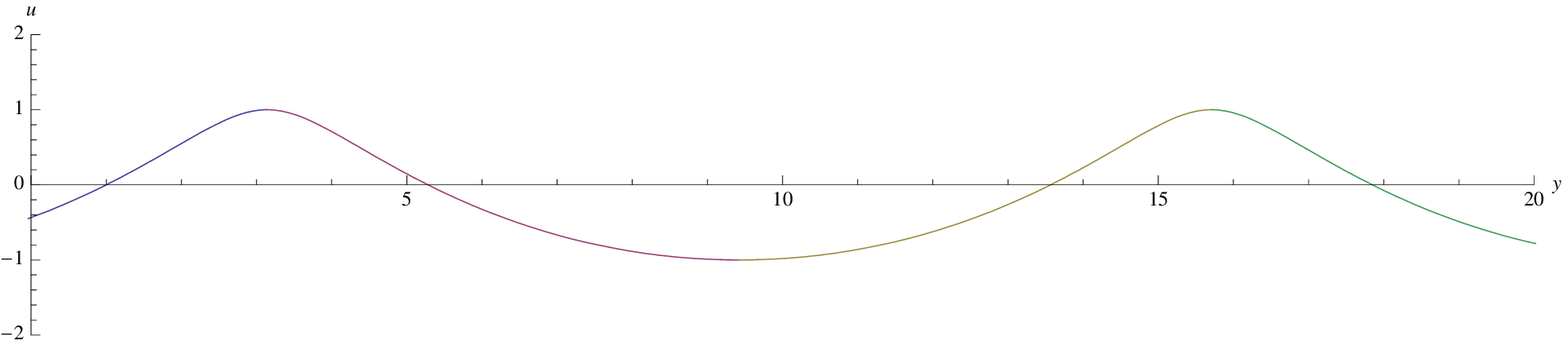} \quad
    \includegraphics[width=.90\textwidth]{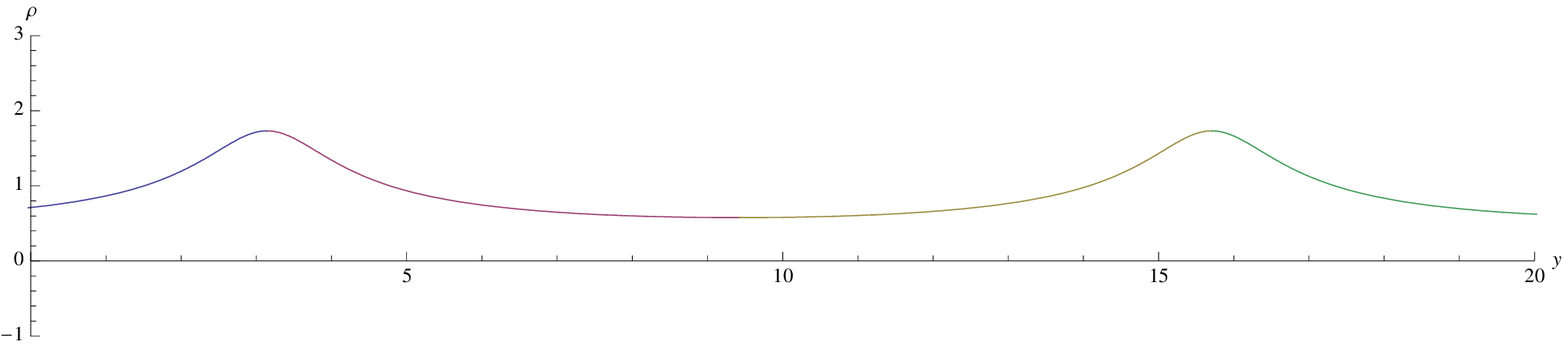} 
     \\
     \begin{figuretext}\label{urhofig}
       A traveling-wave solution $(u, \rho)$ of the two-component Hunter-Saxton equation (\ref{2CHHS}) with $\sigma = 1$ and the parameter values $b = 1$, $\umin = -1$, $\umax=1$, $c= 2$. The top and bottom graphs show $u$ and $\rho$ as functions of $y = x - ct$, respectively.
     \end{figuretext}
     \end{center}
\end{figure}
We can summarize our discussion as follows. Consider the supersymmetric Hunter-Saxton equation for $\sigma =1$ with fields taking values in the Grassmann algebra $\{1, \tau\}$:
\begin{align}\label{HS1tausigma1}
  M_t =& -(M U )_x + \frac{1}{2}\left[(D_1M) (D_1U) + (D_2M) (D_2U)\right],
  	\\ \nonumber
  U =& u + \theta_1 \tau f_1 + \theta_2 \tau f_2 + \theta_2 \theta_1 \rho, \qquad M = iD_1D_2 U.
\end{align}  
For any $\alpha = \pm 1$, $f_0 \in \C$, and $b, \umin, \umax, c \in \R$ satisfying (\ref{bzZcineq}), equation (\ref{HS1tausigma1}) admits the smooth periodic solution
\begin{align} \label{utildeu}
u(x,t) =& \tilde{u}(y)\quad  \text{where $\tilde{u}$ solves}\quad  \tilde{u}_{y}^2 = \frac{b(\umax-\tilde{u})(\tilde{u} - \umin)}{(c - \tilde{u})^2}, \qquad y = x - ct,
	\\ \label{rhopmfrac}
\rho(x,t) =& \alpha \frac{\sqrt{b(c-\umin)(c-\umax)}}{|c - u(x, t)|},
	\\
f_1(x,t) =& if_2(x,t) = f_0 \sqrt{\frac{c-\umin}{c-u}} e^{\alpha\, \text{sgn}(u_x) \frac{i}{2} \arctan\left(\frac{2 \sqrt{(c-\umin)(c-\umax)(\umax-u)(u-\umin)}}{-2 \umin \umax+c (\umin+\umax-2 u)+(\umin+\umax) u}\right)},
\end{align}
where we have suppressed the $(x,t)$-dependence of $u$ in the last equation. In particular, restriction to the bosonic sector yields that $(u, \rho)$ given by (\ref{utildeu})-(\ref{rhopmfrac}) constitute a traveling-wave solution of the two-component Hunter-Saxton equation (\ref{2CHHS}) with $\sigma = 1$, see Figure \ref{urhofig}.

Let us point out that although we cannot present an explicit formula for $u(x,t)$, integration of the ODE for $\tilde{u}$ shows that $u$ is given implicitly up to a translation in $x$ by
$$x - ct= \pm \frac{(u-\umin) (\umax - u)+ \sqrt{(u-\umin)(\umax-u)} (2 c-\umin-\umax) \tan^{-1}\left(\sqrt{\frac{u-\umin}{\umax - u}}\right)}{\sqrt{b (u-\umin) (\umax-u)}}.$$
The stability of these traveling-wave solutions is of interest, especially since in the case of the Camassa-Holm equation it is known that the smooth and peaked traveling waves are orbitally stable cf. \cite{C-Mo2, C-S, C-S2, Lpeakstab}.

\section{Explicit solutions: Second deconstruction} \nequation
In this section we extend the analysis of the previous section to the case of the second deconstruction, that is, to the case of fields taking values in the Grassmann algebra $\{1, \tau_1, \tau_2, \tau_1 \tau_2\}$ where $\tau_1$ and $\tau_2$ are odd variables. Under this assumption we may write
\begin{align}\label{fieldstau1tau2}
  u = u_1 + \tau_2 \tau_1 u_2, \quad \rho = \rho_1 + \tau_2 \tau_1 \rho_2, \quad \varphi_1 = \tau_1 f_1 + \tau_2 g_1, \quad \varphi_2 =  \tau_1 f_2 + \tau_2 g_2,
\end{align}
where $u_1, u_2, \rho_1, \rho_2, f_1, f_2, g_1, g_2$ are real-valued functions of $x$ and $t$. For simplicity we henceforth consider only the case of HS with $\sigma = 1$. In view of the expressions in (\ref{fieldstau1tau2}) for $u, \rho, \varphi_1, \varphi_2$, we find that equation (\ref{superCHHS}) is equivalent to the following list of equations:
\begin{align} \label{e1}
\rho_{1t} + (\rho_1u_1)_x  & = 0,
   	\\ \label{e2}
 -\rho_1 \rho_{1x} + 2u_{1x} u_{1xx} + u_{1txx} + u_1u_{1xxx} & = 0,
   	\\ \label{e3}
 \left(f_1 \rho_1 +2 f_{2t} + 2u_1 f_{2x} + f_2 u_{1x} \right)_x & = 0,
   	\\ \label{e4}
 \left(g_1 \rho_1 + 2 g_{2t} + 2 u_1 g_{2x} + g_2 u_{1x} \right)_x & = 0,
   	\\ \label{e5}
 \left(-f_2 \rho_1 + 2 f_{1t} + 2u_1f_{1x} + f_1 u_{1x} \right)_x & = 0,
   	\\ \label{e6}
 \left(-g_2 \rho_1 + 2 g_{1t} + 2u_1 g_{1x} + g_1 u_{1x} \right)_x & = 0,
   	\\ \label{e7}
2 \rho_{2t} + \left(f_2 g_1 - f_1 g_2 + 2 u_2 \rho_1 + 2 u_1 \rho_2 \right)_x & = 0,
	\\ \nonumber
\bigl( 2 u_{2tx} + g_1f_{1x} + g_2 f_{2x} - f_1 g_{1x} - f_2g_{2x} \qquad \qquad &
	\\ \label{e8}
+ 2 u_{1x} u_{2x} 
 + 2 u_2 u_{1xx} + 2 u_1 u_{2xx} -2 \rho_1 \rho_2 \bigr)_x & = 0.				
\end{align}
Equations (\ref{e1})-(\ref{e8}) ascertain the equality of the coefficients of $1$, $\theta_1 \theta_2$, $\theta_1 \tau_1$, $\theta_1 \tau_2$, $\theta_2 \tau_1$, $\theta_2 \tau_2$, $\tau_1 \tau_2$, $\theta_1\theta_2\tau_1\tau_2$, respectively, on the left- and right-hand sides of equation (\ref{superCHHS}).
We make the following observations:

\begin{itemize}
\item Equations (\ref{e1}) and (\ref{e2}) make up the two-component HS equation (\ref{2CHHS}) for $\sigma = 1$ with $(u, \rho)$ replaced by $(u_1, \rho_1)$. 

\item Equations (\ref{e3}) and (\ref{e5}) are the fermionic equations (\ref{travsystems1}) encountered in the case of the first deconstruction with $(u, \rho)$ replaced by $(u_1, \rho_1)$. 

\item Equations (\ref{e4}) and (\ref{e6}) are the fermionic equations (\ref{travsystems1}) encountered in the case of the first deconstruction with $(u, \rho)$ replaced by $(u_1, \rho_1)$ and $(f_1, f_2)$ replaced by $(g_1, g_2)$.

\item Equations (\ref{e7}) and (\ref{e8}) are the only equations involving $u_2$ and $\rho_2$.
\end{itemize}

In view of these observations, $(u_1, \rho_1, f_1, f_2)$ and $(u_1, \rho_1, g_1, g_2)$ constitute two sets of solutions to the equations considered in the case of the first deconstruction. We can therefore apply the analysis of Section \ref{explicitfirstsec} to obtain explicit expressions for these fields which fulfill equations (\ref{e1})-(\ref{e6}); equations (\ref{e7}) and (\ref{e8}) can then be used to determine $u_2$ and $\rho_2$. Note that for solutions of the form $f := f_1 = i f_2$ and $g := g_1 = i g_2$, equations (\ref{e7}) and (\ref{e8}) reduce to
\begin{align}
    \label{e7red}
 \rho_{2t} + \left(u_2 \rho_1 +  u_1 \rho_2 \right)_x & = 0,
	\\ \label{e8red}
\bigl(u_{2tx} + u_{1x} u_{2x} +  u_2 u_{1xx} + u_1 u_{2xx} -\rho_1 \rho_2 \bigr)_x & = 0.		\end{align}
Assuming that $(u_1, \rho_1)$ is a traveling-wave solution of (\ref{2CHHS}) of the form constructed in Section \ref{explicitfirstsec}, we can obtain solutions $u_2 = u_2(y)$ and $\rho_2 = \rho_2(y)$, $y = x - ct$, of equations (\ref{e7red}) and (\ref{e8red}) as follows. In view of (\ref{travrho}), we have $\rho_1 = a/(c - u_1)$ for some constant $a$. Hence equation (\ref{e7red}) yields
$$ \rho_{2} = \frac{a u_2}{(c - u_1)^2} - \frac{E}{c - u_1},$$
where the constant $E$ arose from an integration with respect to $y$.
Substituing this expression for $\rho_2$ into equation (\ref{e8red}), we find
$$\left[-cu_{2yy} + u_{1y} u_{2y} +  u_2 u_{1yy} + u_1 u_{2yy} - \frac{a^2 u_2}{(c - u_1)^3} + \frac{Ea}{(c - u_1)^2} \right]_x = 0.	$$
Letting $u_2 = c - u_1$, this equation becomes
$$\left[cu_{1yy} - \frac{1}{2}u_{1y}^2 - u_1 u_{1yy} - \frac{a(a-E)}{2(c - u_1)^2}  \right]_x = 0.$$
A comparison with (\ref{cuyytrav}) shows that this equation is fulfilled provided that $E = 2a$. We conclude that (\ref{e7red}) and (\ref{e8red}) are satisfied for
$$u_2 = c - u_1, \qquad \rho_2 = - \frac{a}{c - u_1}.$$
We note that $\rho_2 = - \rho_1$ for this solution.

\subsection{Summarized result}
The discussion in this section can be summarized as follows. Consider the supersymmetric Hunter-Saxton equation for $\sigma =1$ with fields taking values in the Grassmann algebra $\{1, \tau_1, \tau_2, \tau_1\tau_2\}$:
\begin{align}\label{HStau1tau2sigma1}
  M_t =& -(M U )_x + \frac{1}{2}\left[(D_1M) (D_1U) + (D_2M) (D_2U)\right],
  	\\ \nonumber
  U =& (u_1 + \tau_2 \tau_1 u_2) + \theta_1 (\tau_1 f_1 + \tau_2 g_1) + \theta_2 (\tau_1 f_2 + \tau_2 g_2) + \theta_2 \theta_1 (\rho_1 + \tau_2 \tau_1 \rho_2), 
  	\\ \nonumber
   M =& iD_1D_2 U.
\end{align}  
For any $\alpha = \pm 1$, $f_0, g_0 \in \C$, and $b, \umin, \umax, c \in \R$ satisfying (\ref{bzZcineq}), equation (\ref{HStau1tau2sigma1}) admits the smooth periodic solution
\begin{align*}
u_1(x,t) =& \tilde{u}(y)\quad  \text{where $\tilde{u}$ solves}\quad  \tilde{u}_{y}^2 = \frac{b(\umax-\tilde{u})(\tilde{u} - \umin)}{(c - \tilde{u})^2}, \qquad y = x - ct,
	\\ \label{rhopmfrac}
\rho_1(x,t) =& \alpha \frac{\sqrt{b(c-\umin)(c-\umax)}}{|c - u_1(x, t)|},
	\\
f_1(x,t) =& if_2(x,t) = f_0 \sqrt{\frac{c-\umin}{c-u_1}} e^{\alpha\, \text{sgn}(u_{1x}) \frac{i}{2} \arctan\left(\frac{2 \sqrt{(c-\umin)(c-\umax)(\umax-u_1)(u_1-\umin)}}{-2 \umin \umax+c (\umin+\umax-2 u_1)+(\umin+\umax) u_1}\right)},
	\\
g_1(x,t) =& ig_2(x,t) = g_0 \sqrt{\frac{c-\umin}{c-u_1}} e^{\alpha\, \text{sgn}(u_{1x}) \frac{i}{2} \arctan\left(\frac{2 \sqrt{(c-\umin)(c-\umax)(\umax-u_1)(u_1-\umin)}}{-2 \umin \umax+c (\umin+\umax-2 u_1)+(\umin+\umax) u_1}\right)},
	\\
u_2(x,t) =& c - u_1(x,t), 
	\\
\rho_2(x,t) =& - \rho_1(x,t).
\end{align*}

 \bigskip
\noindent
{\bf Acknowledgement} {\it J.L. acknowledges support from a Marie Curie Intra-European Fellowship. The work of O.L. is partially supported by DFG.}

\bibliography{is}

\end{document}